\documentclass[showpacs,amsmath,amssymb]{revtex4}

\usepackage{graphicx}    
\usepackage{dcolumn}     
\usepackage{bm}          

\begin{document}


\title{ADC non-linear errors correction in thermal diagnostics for the
LISA mission}

\author{J. Sanju\'an}
\author{A. Lobo}
\affiliation{%
Institut de Ci\`encies de l'Espai (CSIC-IEEC), \\ Edifici Nexus,
Gran Capit\`a 2-4, 08034 Barcelona, Spain} 
%
%
\author{J. Ramos-Castro}
\affiliation{Departament d'Enginyeria Electr\`onica, Universitat
Polit\`ecnica de Catalunya (UPC), Edif. C4, Jordi Girona 1-3,
08034 Barcelona, Spain}%


\begin{abstract}
Low-noise temperature measurements at frequencies in the milli-Hertz range 
are required in the LISA (Laser Interferometer Space Antenna) and LISA
PathFinder (LPF) missions. The required temperature stability for LISA
is around 10\,$\mu$K\,Hz$^{-1/2}$ at frequencies down to 0.1\,mHz. In
this paper we focus on the identification and reduction of a source of
excess noise detected when measuring time-varying temperature signals.
This is shown to be due to non-idealities in the ADC transfer curve,
and degrades the measurement by about one order of magnitude in the
measurement bandwidth when the measured temperature exhibits drifts
of $\sim\mu$K\,s$^{-1}$. In a suitable measuring system for the LISA
mission, this noise needs to be reduced. Two different methods based
on the same technique have been implemented, both consisting in the
addition of dither signals out of band to mitigate the ADC non-ideality
errors. Excess noise of this nature has been satisfactorily reduced by
using these methods when measuring temperature ramps up to 10~$\mu$K\,s$^{-1}$.
\end{abstract}


\pacs{07.20.Dt, 07.87.+v}

\maketitle

\section{Introduction}
\label{intro}

\begin{figure}[h]
 \includegraphics[width=0.65\columnwidth]{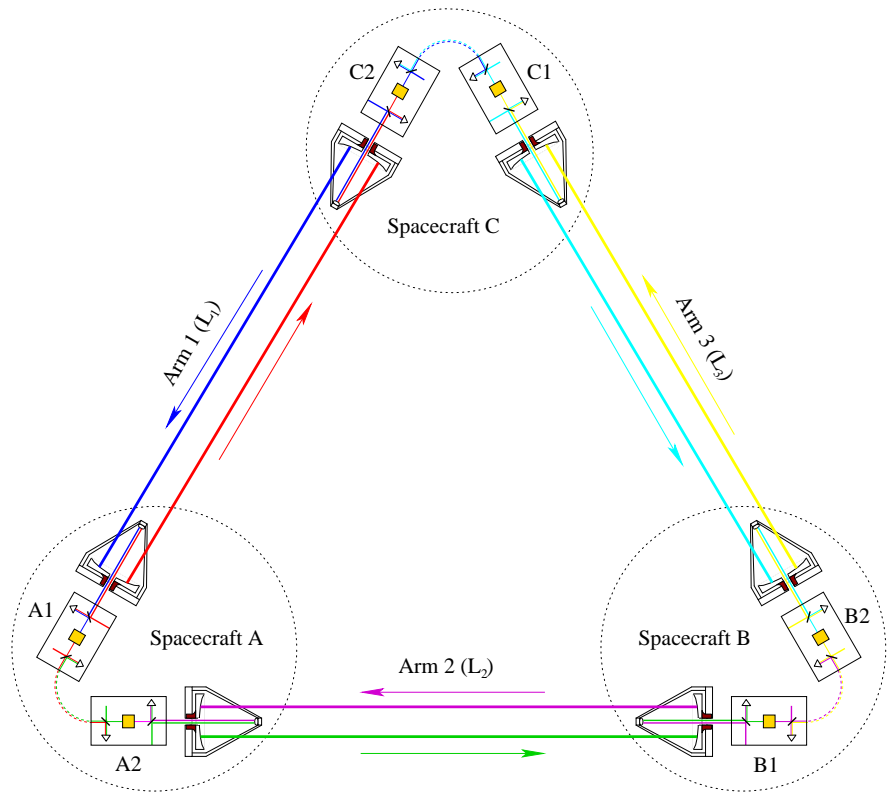}
 \caption{Conceptual drawing of the LISA constellation. Arm lengths are
 nominally equal, and each is 5 million km long (graph is not to scale).
\label{fig0}}
\end{figure}

LISA (Laser Interferometer Space Antenna) is a joint ESA-NASA space mission
conceived to detect Gravitational Waves (GWs)~\cite{bender}. LISA consists
in a constellation of three spacecraft in the vertexes of an equilateral
triangle 5~million kilometres to the side. The constellation orbits the
Sun following the ecliptic, some 20~degrees (45~million kilometres)
behind the Earth. Each spacecraft houses two proof masses in nominal
free fall, and laser links are established between spacecraft ---see
Figure~\ref{fig0}. The role of the links is to enable interferometric
measurements of relative distance and acceleration variations between
pairs of proof masses in distant spacecraft. This is how LISA will
detect GWs, since they show up as \emph{tidal forces} (or \emph{geodesic
deviations} in the language of General Relativity Theory) in the region
where the detector is.

Expected GW signals are extremely weak~\cite{sources}, hence severe
requirements must be imposed on the proof mass residual acceleration
noise to ensure LISA meets its GW detection objectives. These are
defined in terms of spectral density~\cite{lisasrd}:
\begin{equation}
 S_{\delta a, {\rm LISA}}^{1/2}(\omega)\leq 3\times 10^{-15}\cdot
 \left\{\left[
 1 + \left(\frac{\omega/2\pi}{8\ {\rm mHz}}\right)^{\!\!4}\right]\!
 \left[1 + \left(\frac{0.1\ {\rm mHz}}{\omega/2\pi}\right)\right]
 \right\}^{\!\frac{1}{2}} \ {\rm m}\,{\rm s}^{-2}\,{\rm Hz}^{-1/2} \nonumber
 \label{lisa.req.acc}
\end{equation}
in the frequency band from 0.1~mHz to 0.1~Hz, which is where optimum GW
detection performance can be obtained with interferometer arm-lengths of
$\sim$5~million kilometres~\cite{lobo.cqg92}.

The conditions in Equation~(\ref{lisa.req.acc}) are not only very demanding,
they cannot be directly put to test in an earth based environment. The reason
is of course the impossibility to maintain a physical system in accurate free
fall near the Earth's surface during periods of several hours, as required
for measurements in sub-milli-Hertz frequency bands. This fact motivated ESA,
the European Space Agency, to fly a technology precursor mission to secure
proper working of the key technologies needed by LISA. Such technology
mission goes by the name LISA PathFinder (LPF), and its launch date is
scheduled for 2011. LPF is a squeezed version of one LISA arm: its length
is downscaled to $\sim$30\,cm, and the two proof masses are housed in a
single spacecraft. Noise requirements in LPF are relaxed by about an order
of magnitude relative to those of LISA. Also, the Measuring Bandwidth (MBW)
in the LTP (LISA Technology Package, the main instrument on board LPF) is
reduced to the range between 1~mHz and 30~mHz. Because of its small
dimensions, LPF cannot work as a milli-Hertz GW detector; rather, LPF is a
noise monitor intended to understand and properly model its sources~\cite{anza}.

One of such sources of noise, which can limit the performance of LISA, is
temperature fluctuations. Thermal stability is necessary to ensure the
stability of the optical elements properties in the Optical Metrology
System (Optical Bench, mirrors, beam-splitters, etc.), as well as of
the proof masses' environment, where such effects as radiometer and/or
radiation pressure fluctuations will cause random accelerations if the
temperature fluctuates. This happens in LISA and LPF alike, and the
maximum tolerable levels of temperature fluctuations can be derived
from estimates of the magnitude of their contribution to the total
instrument noise, and (conventionally) assuming that they should
contribute not more than $\sim$10\,\% to the total. This generates
a requirement in temperature stability for LISA of~\cite{bender}
\begin{equation}
  S^{1/2}_{T,\, \rm  LISA}(\omega)\lesssim 10\,\mu{\rm K}\,{\rm Hz}^{-1/2}
  \label{eq.1}
\end{equation}
in LISA's MBW, i.e., from 0.1~mHz to 0.1~Hz. Consequently, measurements
capable to discern such small temperature fluctuations are needed. In LPF,
the requirements being relaxed as described above, the temperature stability
is also less severe~\cite{on.ground.tests}:
\begin{equation}
  S^{1/2}_{T,\, \rm  LTP}(\omega)\leq 100\,\mu{\rm K}\,{\rm Hz}^{-1/2}
  \label{eq.2}
\end{equation}
in the frequency band from 1~mHz to 30~mHz. In order to make meaningful
temperature measurements, the measuring system must be quieter than
this figure, and again 10\,\% of the above is required for it, i.e.,
\begin{equation}
  S^{1/2}_{T,\,\rm  system}(\omega)\leq 10\,\mu{\rm K}\,{\rm Hz}^{-1/2}
  \label{eq.3}
\end{equation}

The LTP temperature monitoring system has already been designed and
tested~\cite{fee.paper} ---see also Section~\ref{fee.description} below---
and it works. However, a small excess noise has been detected at frequencies
around the milli-Hertz when the measured temperature drifts with time.
This effect does not pose a serious problem in the LTP, as it only shows
up at very low frequencies ($<1$\,mHz) and in the presence of somewhat
high temperature drifts, $\left|dT/dt\right|$\,$\sim\mu$K\,s$^{-1}$. In LISA
the MBW stretches down to 0.1~mHz~\cite{feedback.paper}, and even weaker
temperature drifts are likely to significantly deteriorate the performance
of the measuring system. As we shall see, its origin has been traced to
non-linearities of the Analog-to-Digital Converters (ADC).

This paper focuses on the identification and reduction of this excess noise.
It is organised as  follows: Section~\ref{fee.description} briefly describes
the temperature measurement system which will actually fly in LPF.
Section~\ref{problem} delves into the details of a potential problem
with the non-idealities of the ADCs. In Section~\ref{mitigation} we
present experimental evidence that the observed excess noise can indeed
be attributed to the ADCs non-ideal response, and also two different ways
to reduce that noise are analysed. In Section~\ref{results} we give details
of the experimental setups, tests carried through to assess the quality
of the proposed methods, and results. Finally, Section~\ref{discussion}
highlights the most relevant results and the prospects of their
applicability for LISA.

\section{Temperature measurement system description}
\label{fee.description}

\begin{figure}[h]
  \includegraphics[width=0.65\columnwidth]{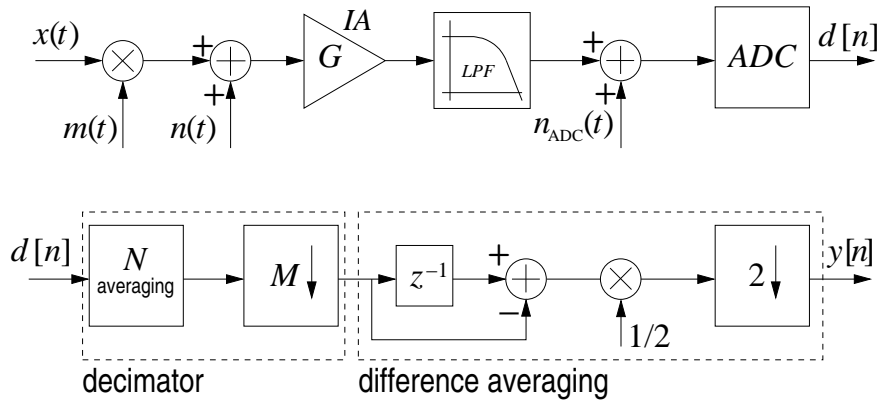}
  \caption{Temperature measurement system chain. \label{fee.proc}}
\end{figure}

In this section the temperature measurement system is briefly reviewed.
Details can be found in~\cite{fee.paper}. The functional block diagram 
of the system is given in Fig.~\ref{fee.proc}.

The temperature signal, $x(t)$, coming from a Wheatstone bridge is
modulated by a square wave signal of angular frequency $\omega_{\rm c}$
($\omega_{\rm c}/2\pi$\,=\,6.25\,Hz) to avoid the 1/$f$ noise of the
Instrumentation Amplifier (IA). The modulated signal is amplified by
the IA and low-pass filtered to avoid aliasing. The signal is then
quantised by a 16-bit ADC. Once the signal is digitised, a digital
demodulation takes place: $N_{\rm av}$ samples are averaged during 
either polarity of the modulating square wave, and the averaged
value of one polarity is subtracted from that of the other one
---see schematics in Fig.~\ref{fee.proc}. This process results
in a noise equivalent temperature for \emph{dc} signals of
\begin{equation}
  S_{T}^{1/2}(\omega)\approx 10\,\mu{\rm K}\,{\rm Hz}^{-1/2} 
  \label{eq.4}
\end{equation}
at frequencies down to 0.1~mHz~\cite{feedback.paper,fee.paper}.

The signal coming from the analog circuit of the front-end electronics (FEE)
is quantised by an ADC. Only one type of ADC suitable for this purpose has
been certified for use in space applications; it is a 16-bit capacitor
based Successive Approximation Register (SAR). Limitations in the
performance of the system at low frequencies (milli-Hertz range) when slowly
drifting input signals ($\sim\mu$V\,s$^{-1}$) are quantised are related to
the non-idealities of such ADCs, as will be shown below.

\section{Non-ideal quantisation noise}
\label{problem}

In this section we analyse the ADC non-linearity errors, and how
they perturb the temperature measurements.

Quantisation is inherently a non-linear process. By construction, the
values of the analog function are rounded up to the closest ADC step,
so that even an ideal ADC generates output errors associated to the
differences between the real and quantised values of the signal. In an
ADC~\cite{jespers} the step size, or Least Significant Bit (LSB),
is defined by
\begin{equation}
 \Delta = \frac{1}{2^{N}-1}V_{\rm FS}
 \label{eq.5}
\end{equation}
where $N$ is the number of bits of the ADC, and $V_{\rm FS}$ is the maximum
voltage (full-scale) the ADC can quantise. When a large number of bits is
considered the error introduced by the
quantisation process is usually treated as an independent random variable
with uniform probability density function (pdf) and white spectral
density~\cite{jespers}.

\begin{figure}[h]
 \includegraphics[width=0.5\columnwidth]{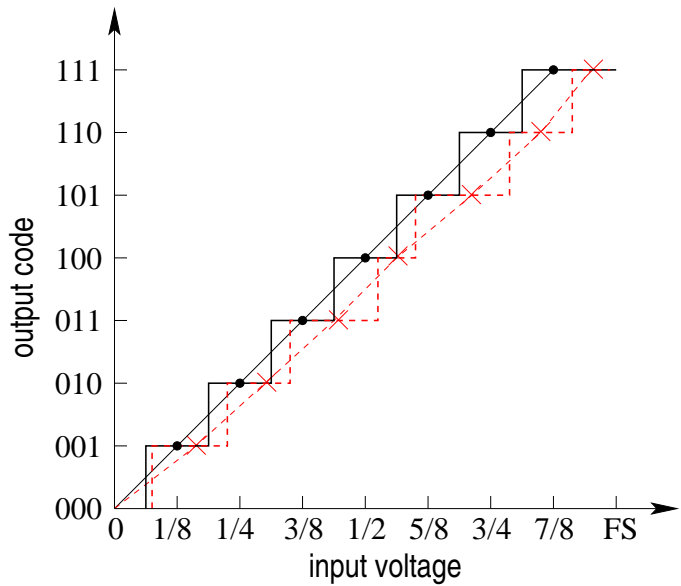}
 \caption{INL error effect on a 8-bit ADC: the INL is the difference
  between dots (ideal ADC) and crosses (real ADC).
 \label{nonlinear}}
\end{figure}

In a real ADC, however, the quantisation steps are not uniform due to
mismatches in the internal topology of the ADC; more specifically,
tolerances in the capacitors of the SAR
array~\cite{hoeschele,analog.dev.book,arpaia} ---see Figure~\ref{nonlinear}.
Such non-uniformity is specified with two parameters: Differential
Non-Linearity (DNL) and Integral Non-Linearity (INL) errors. 
The DNL is defined as the deviation between two adjacent transition
points of the quantisation ladder and an LSB, or~\cite{Michaeli2008,Marc1997}
\begin{equation}
 DNL(i) = K_{i+1}-K_{i}-\Delta, \qquad i=0,1,2,...,2^{N}-1
 \label{eq.6}
\end{equation}
where $K_i$ is the output value of the $i\/$-th quantisation code.

The additional noise related to the DNL errors can be reduced by suitable
dithering~\cite{carbone94, gray93, sch, wagdy96}. The inherent ADC noise
plus FEE noise can be considered in this case as a dither source, which
appears to be enough to make the DNL negligible in practice ---see
Section~\ref{real.adcs} below.

On the other hand, the INL error is defined as the discrete integral of
the DNL, i.e.,
\begin{equation}
 INL(i) = \sum^{i}_{j=0}DNL(j), \qquad i=0,1,2,...,2^{N}-1
 \label{eq.7}
\end{equation}
and can be understood as the difference between the real and the ideal
ADC transfer curves ---see Figure~\ref{nonlinear}. The noise introduced
by this error is usually less noticeable, although much more difficult
to reduce. In this paper we focus on how to do this, since the INL is
the one limiting the performance of the measurement when slowly drifting
input signals are present. This is a common situation in LTP temperature
measurements.

Experimental results ---see Section~\ref{non.stationary}--- obtained with
the temperature measurement system reflect the problem depicted in
Figure~\ref{nonlinear}, which shows up as excess noise in the spectrum.
Such noise could not be
attributed to actual temperature fluctuations since sensors were placed
in a thermal environment where fluctuations were efficiently screened
out~\cite{on.ground.tests}, hence such behaviour may not be due to variable
temperatures. Analog noise or interferences coming from the signal
processing chain (Wheatstone bridge, amplification, low-pas filtering)
were also discarded as possible sources of the extra noise. Finally,
the excess noise disappears when a \emph{dc} temperature
value (i.e., not drifting with time) is measured~\footnote{
For instance, using a high-stability resistor instead of a sensor.}
---see Figure~\ref{spec1}. All in all, the origin of the excess noise 
has been identified as due to INL errors of the ADC. In the following
we present more quantitative arguments confirming this hypothesis.

\subsection{ADC bit error description}

We first analyse the effect of a faulty bit in a SAR ADC on the performance
of the measurement. For this, a simple ADC model is considered. The
analog-to-digital conversion is done (in SAR ADCs) by comparing the
sampled signal with an analog voltage generated by a Digital-to-Analog
Converter (DAC) and a SAR~\cite{hoeschele}. The topology of the DAC is
based on a switching capacitor bank composed by 16 capacitors, scaled
from $2^{16}C$ to $C$, where $2^{16}C\/$ defines the Most Significant
Bit (MSB) and $C\/$ defines the LSB. The DAC output voltage is~\cite{jespers}
\begin{equation}
 V_{\rm DAC}=\frac{\sum^{N-1}_{k=0}b_k 2^k C}{\sum^{N-1}_{k=0}2^k C}\,
 V_{\rm FS} = \frac{\sum^{N-1}_{k=0}b_k 2^k C}{(2^{N}-1)\,C}\,V_{\rm FS}
 \label{eq.8}
\end{equation}
where $b_{k}$ is the binary digit ($k=0$ stands for the LSB and $k=N-1$
for the MSB); it is set to 0 or 1 depending on the sampled voltage, and
$N\/$ is the number of bits of the ADC. An error
$\delta C_k$\,=\,$C_{k,{\rm real}}$$-$$2^k C$ in the $k\/$-th capacitor
results in an error in the DAC output when the corresponding bit is set
to 1. The voltage error $\epsilon_k$ induced by a faulty $k\/$-th bit is
therefore
\begin{equation}
 \epsilon_k\simeq\frac{b_k\,\delta C_k}{(2^N-1)\,C}\,V_{\rm FS}
 = b_k\,\frac{\delta C_k}{C}\,\Delta
 \label{epsilon}
\end{equation}
where $\Delta$ is the ideal LSB.

\begin{figure}[h]
 \includegraphics[width=0.6\columnwidth]{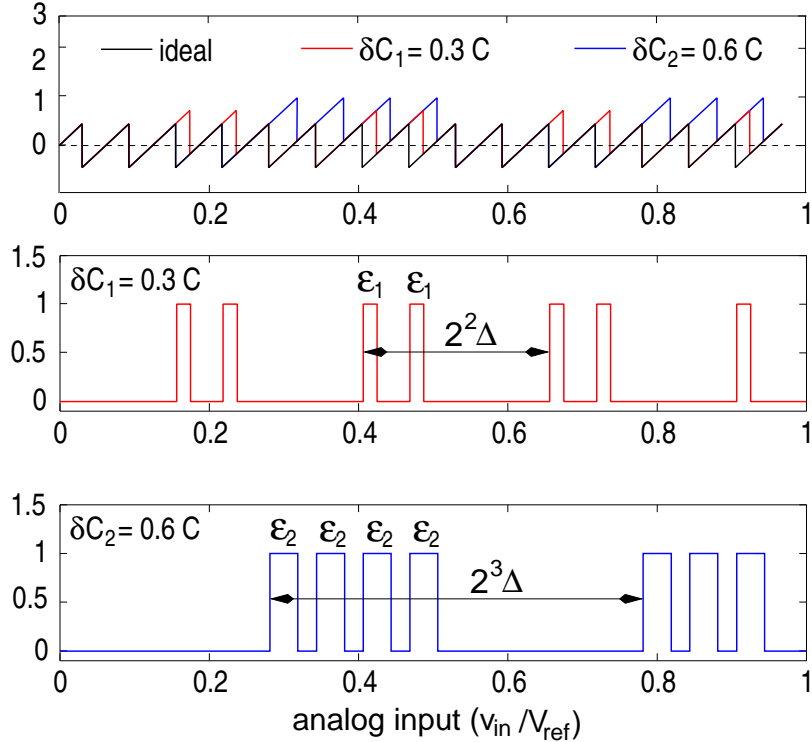}
 \caption{Three simulated 4-bit ADCs. Top: quantisation error functions
 for the ideal ADC (black trace), a real ADC with a faulty $k\/$\,=\,1~bit
 ($\epsilon_{1}=0.3\Delta$, red), and a real ADC with a faulty $k\/$\,=\,3
 bit ($\epsilon_{1}=0.6\Delta$, blue). Centre and bottom: error differences
 between the two real ADCs and the ideal one, respectively. Units in
 ordinates are LSBs.
 \label{sawtooth}}
\end{figure}

The erroneous bit produces a superimposed periodic pattern in the
quantisation error of the ADC with 1 LSB of amplitude when the input
voltage varies between 0 and $V_{\rm FS}$, and exhibits a periodicity with
the ADC input voltage of $2^{k+1}\Delta$. An example of this is shown in
Figure~\ref{sawtooth} where it can be seen how the faulty bit introduces
a long period component superimposed on the typical sawtooth error
function of an ideal ADC.

\subsection{Dither signal effect in ideal quantisers}
\label{dither.section}

When a dither voltage, $d$, with a certain probability density function
(pdf), $p(d)$, is added to the input signal of the ADC, $v$, the average
quantisation error observed at the ADC output, $\langle q(v)\rangle$, is
given by~\cite{wagdy89c}
\begin{equation}
 \langle q(v)\rangle = \int^{\infty}_{-\infty}\,q(v+z)\,p(z)\,dz
 \label{eq.9}
\end{equation}
where $q(v)$ is the quantisation error of an ideal ADC. If we define the
Fourier transform in voltage domain by \footnote{
$\xi$ is the Fourier-conjugate variable of $v\/$, and is accordingly
measured in $2\pi$\,volts$^{-1}$.}
\begin{equation}
 {\cal Q}(\xi)\equiv\int_{-\infty}^\infty\, q(v)\,e^{-i\xi v}\,dv
 \label{eq.10}
\end{equation}
and apply it to Eq.(\ref{eq.9}), the following ensues:
\begin{equation}
 \langle {\cal Q}(\xi)\rangle = {\cal Q}(\xi)\,{\cal P}^*(\xi)
 \label{dith.1}
\end{equation}

The quantisation error for an ideal ADC (a sawtooth waveform of amplitude
$\Delta$/2 ---see Figure~\ref{sawtooth}) is a periodic function of $v\/$
with period $\Delta$. It can be expanded in Fourier series:
\begin{equation}
  q(v)=\frac{\Delta}{\pi}\sum^{\infty}_{n=1}\frac{(-1)^{n+1}}{n}\,
 \sin\frac{2\pi nv}{\Delta}
 \label{qideal}
\end{equation}
and its Fourier transform is
\begin{equation}
 {\cal Q}(\xi) = -i\,\Delta\sum_{\scriptsize\begin{array}{c}
 n=-\infty \\ (n\neq 0) \end{array}}^{\infty}\,
 \frac{(-1)^{n+1}}{n}\;\delta\!\left(\xi-\frac{2\pi n}{\Delta}\right)
 \label{Qwx}
\end{equation}
where $\delta(\cdots)$ is Dirac's $\delta$-function. If we consider
zero-mean Gaussian noise as the dither voltage, its pdf and its Fourier
transform are, respectively,
\begin{eqnarray}
 p(d) & = & \frac{1}{\sqrt{2\pi\sigma^2}}\,e^{-\frac{d^2}{2\sigma^2}} \\[0.7ex]
 {\cal P}(\xi) & = & e^{-\sigma^2\xi^2/2}
 \label{dith.fourier}
\end{eqnarray}
where $\sigma^2$ is the dither variance. Substituting now Eqs.~(\ref{Qwx})
and~(\ref{dith.fourier}) into Eq.~(\ref{dith.1}) and taking the modulus,
we obtain
\begin{equation}
 |\langle {\cal Q}(\xi)\rangle| = \Delta\cdot e^{-\sigma^2\xi^2/2}\,
 \sum^{\infty}_{n=-\infty \atop (n\neq0)}\,\frac{1}{n}\;
 \delta\!\left(\xi-\frac{2\pi n}{\Delta}\right)
 \label{dith.eq}
\end{equation}

From Eq.~(\ref{dith.eq}) we note that the Gaussian dither signal, $d$,
low-pass filters the quantisation error ---see Figure~\ref{dither1}, top.
For instance, for $\sigma$\,=\,$\Delta$ the attenuation of the first term
in the series is 2.7$\times$$10^{-9}$. In our case, we have a 16-bit ADC
with $V_{\rm FS}$=10\,V, \emph{quasi}-white noise at the input of the ADC
of $S^{1/2}_{V}\simeq 7$\,$\mu$V\,Hz$^{-1/2}$ and a noise-equivalent
bandwidth (NEBW) of 1.2\,$f_{\rm cut-off}$\,=\,600\,Hz, where $f_{\rm cut-off}$
is the cut-off frequency of the anti-alias filter~\cite{fee.paper}. Thus,
$\Delta$ and $\sigma$ are readily calculated, i.e.,
\begin{eqnarray}
 \Delta & = & \frac{V_{\rm FS}}{2^N-1} = 0.15\,{\rm mV} \\[1ex]
 \sigma & = & S^{1/2}_{V}(\omega)\times(1.2\,f_{\rm cut-off})^{1/2} =
 0.17\,{\rm mV} 
\end{eqnarray}

In our specific case, $\sigma\simeq\Delta$, hence the quantisation noise
from the ideal ADC should be suppressed by the inherent noise of the analog
processing chain which acts as a filter to the ideal quantisation error.
This can be very clearly seen in Figure~\ref{dither1}, top, which shows
the errors of an ideal ADC (vertical lines) alongside the equivalent
low-pass filter profile (dashed trace) achieved by means of Gaussian
dither with $\sigma$\,=\,$\Delta$.

\subsection{Dither signal effect in non-ideal quantisers}
\label{real.adcs}

The error of a real ADC is formed by the ideal quantisation error function
---see Eq.~(\ref{qideal})--- plus a term related to the non-idealities
of the ADC ---see Figure~\ref{sawtooth}. Thus,
\begin{equation}
 q(v)=q_{\rm i}(v) + q_k(v)
\end{equation}
where $q_{\rm i}(v)$ and $q_k(v)$ are the ideal quantisation error and
the quantisation error due to the faulty bits, respectively. As seen in
the previous section, the ideal quantisation error is filtered out by
the analog noise in the measurement chain, which acts as a dither signal
of Gaussian pdf. Instead, the quantisation error due to the non-ideality
of the ADC is not reduced by the same analog noise, which causes it to
show up as extra noise. In this section we show how it can be identified
in the temperature measurements, and how it limits the performance of
the system.

The modulus of the Fourier transform of the non-ideal noise for a single
defective bit, $k$, say, is~\cite{wagdy96}
\begin{equation}
 |{\cal Q}_k(\xi)| = \Delta\sum_{n=-\infty}^\infty\,
 \frac{\sin\frac{n\pi\epsilon_k}{2^{k+1}\Delta}}{n}\,
 \frac{\sin \frac{n\pi}{2}}{\sin \frac{n\pi}{2^{k+1}}}\;
 \delta\!\left(\xi-\frac{\pi n}{2^{k}\Delta}\right)
 \label{qq}
\end{equation}
where $\epsilon_k$ is given in Eq.~(\ref{epsilon}). The effect of
Gaussian dither on $q_k(v)$ is readily calculated with Eq.(\ref{dith.1}):
\begin{equation}
 |\langle{\cal Q}_k(\xi)\rangle| = 
 \Delta \cdot e^{-\xi^2\sigma^2/2}\sum_{n=-\infty}^\infty\,
 \frac{\sin\frac{n\pi\epsilon_k}{2^{k+1}\Delta}}{n}\,
 \frac{\sin \frac{n\pi}{2}}{\sin \frac{n\pi}{2^{k+1}}}\;
 \delta\!\left(\xi-\frac{\pi n}{2^{k}\Delta}\right)
 \label{qqq}
\end{equation}

\begin{figure}[h]
 \includegraphics[width=0.6\columnwidth]{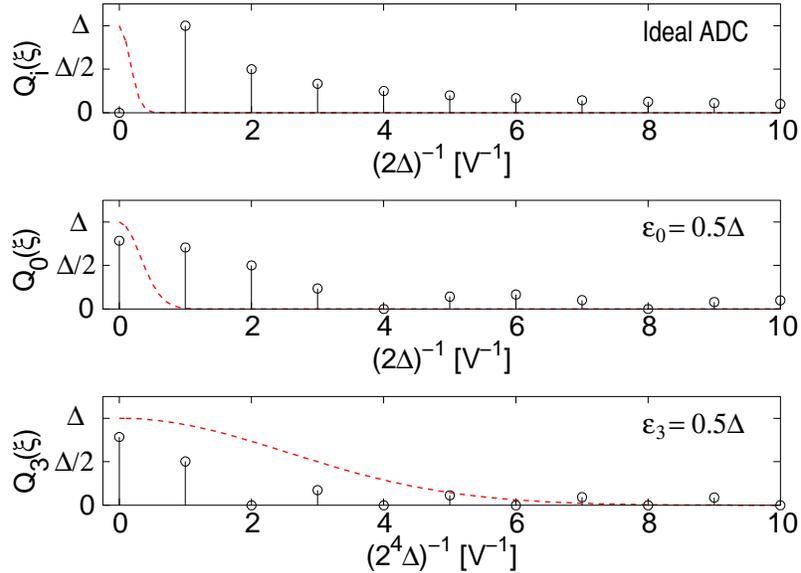}
 \caption{Top: quantisation error for an ideal ADC, ${\cal Q}_{\rm i}(\xi)$.
 Centre: quantisation error for a real ADC with $k\/$\,=\,0 as the faulty
 bit. The error is $\epsilon_0$\,=\,0.5$\Delta$. Bottom: same as previous
 but for $k\/$\,=\,3. The error is $\epsilon_3$\,=\,0.5$\Delta$. Note that
 the $x\/$-axis scale of the bottom plot is different from the top and
 middle ones; for instance, the main frequency for ${\cal Q}_{0}$ is
$1/2\Delta$,
 while that for ${\cal Q}_{3}$ is $1/2^4\Delta$. This means that errors in the
 higher bits show up at lower frequencies than those due to errors in  the LSB.
 \label{dither1}}
\end{figure}

Equations~(\ref{qq}) and~(\ref{qqq}) are plotted in Figure~\ref{dither1}
(centre and bottom). The plot in the centre corresponds to errors associated
to $k\/$\,=\,0, i.e., the LSB. Here, the fundamental period is $2\Delta$.
The plot in the bottom shows the same for $k\/$\,=\,3 (the 4-th bit) with
a fundamental period of 16$\Delta$. For each plot, the low-pass filter
generated by Gaussian dither of $\sigma$\,=\,$\Delta$ is also plotted
(dashed lines). It can be noticed that this dither signal suffices to
suppress the noise associated to a faulty LSB, $k\/$\,=\,0 (this
corresponds to the DNL effect which, is readily mitigated). However,
it cannot attenuate the low frequency lines due to errors in the higher
bits. In fact, when $k\/$\,=\,3 the dithering with $\sigma$\,=\,$\Delta$
is clearly insufficient.

From Figure~\ref{dither1} we note how faulty bits introduce spectral
components in the measurement if the signal spans a large enough fraction
of the range of the ADC transfer curve. Conversely, if the input signal
is a constant \emph{dc} then no extra noise will be seen, since no bit,
whether faulty or not, will change its state. Therefore, the INL effect
becomes perceptible in our system only when the input signal runs through a
sufficiently wide fraction of the ADC range.

\subsection{INL effect on general signals}
\label{non.stationary}

As we have just seen, when the input signal, $v$, is not a \emph{dc}
constant (e.g., a ramp), the errors related to the INL of the ADC tend
to introduce low frequency noise components which degrade the performance
of the system. In fact, this is seen to happen in our device for voltage
variations in the order of few milli-volts. Moreover, if the input signal
changes rate, i.e., $\ddot{v}\neq0$ then the frequency components introduced
by the INL errors spread across the frequency band.

In order to deal with non-constant signals and to estimate INL induced
noise, a simplification is expedient. First, let us assume the input
signal is a straight line for a certain time interval, i.e.,
\begin{equation}
 v(t)= \left\{ \begin{array}{ll} a+bt\ \ &
 {\rm if}\ \ 0\leq t\leq t_{\rm max}\\[1ex]
 0 \ \ & {\rm if}\ \ t\leq 0\ \ {\rm or}\ \ t> t_{\rm max} \end{array}\right.
 \label{eq.11}
\end{equation}
where $a\/$ is the initial value of the signal $v(t)$, and $b\/$ is the
slope of the input signal (V\,s$^{-1}$). If we note $\tilde{q}_{k}(\omega)$
the usual time-frequency Fourier transform of the non-ideal quantisation
noise, it immediately follows from Eq.~(\ref{eq.11}) that
\begin{equation}
 \tilde{q}_{k}(\omega) = \frac{e^{i\omega a/b}}{|b|}\,
 {\cal Q}\left(\frac{\omega}{|b|}\right)
 \label{eq.12}
\end{equation}

In this case, Eq.~(\ref{qq}) becomes
\begin{equation}
 |\tilde{q}_{k}(\omega)|=
 \frac{\Delta}{\pi}\,\sum^{\infty}_{n=-\infty}
\frac{\sin\frac{n\pi\epsilon_{k}}{2^{k+1}\Delta}}{n}\, \frac{\sin \frac{n\pi}{2}}%
 {\sin\frac{n\pi}{2^{k+1}}}\;
 \label{q1}
 \delta\!\left(\omega-\frac{\pi n|b|}{2^{k}\Delta}\right)
\end{equation}
and the main frequency component for an ADC with an error in the
$k\/$-th bit is located at
\begin{equation}
 \omega_{1_k}=\frac{\pi|b|}{2^{k}\Delta}
 \label{w1k}
\end{equation}

Before we proceed further, a technical comment is in order. In several
equations above, Dirac $\delta$-functions appear. They are the result of
infinite length integration intervals in Fourier transform calculations,
which are of course idealisations. In any practical situation, such intervals
are limited to the experimentally available data ranges, so that the
$\delta$-functions are actually sinc-functions: they have identical
centre points but spread around those centres depending on the integration
interval lengths. In order to make sense of e.g.\ Eq.~(\ref{q1}) we must
ask which is the minimum required integration time to obtain a meaningful
spectrum or, in other words, which is the minimum duration, $t_{\rm max}$
of the ramp signal in Eq.~(\ref{eq.11}). This is easily inferred from the
frequency of the lowest spectral line $\omega_{1_k}$ in Eq.~(\ref{w1k}):
for a conventional ten cycle integration time we get
\begin{equation}
 t_{\rm max} > 10\,\frac{2\pi}{\omega_{1_k}} =
 10\,\frac{2^{k+1}\,\Delta}{|b|}
 \label{eq.13}
\end{equation}

These considerations apply to the analysis of the ADC response to signals
which drift slowly with time, where \emph{slow drift} means such signals
can be conveniently approximated by a series of concatenated ramps with
suitable slopes, and lengths complying with Eq.(\ref{eq.13}). In these
circumstances, we can generalise Eq.~(\ref{w1k}) as follows:
\begin{equation}
 \omega_{1,k} = \frac{\pi}{2^k\,\Delta}\,|\dot{v}(t)|\ ,\qquad
 \dot{\omega}_{1,k} = \frac{\pi}{2^k\Delta}\,\ddot{v}(t)
 \label{dw}
\end{equation}

From Eqs.~(\ref{w1k}) and~(\ref{dw}) some conclusions can be drawn:

\begin{itemize}
\setlength{\itemsep}{-0.4 ex}   
 \item High-slope signals at the ADC input translate into high-frequency
 components at the output of the ADC.
 \item Errors in the higher bits show up as noise at low frequency.
 \item Consequently, high-slope input signals combined with errors in high
 bits may show as noise peaks in the measurement bandwidth.
 \item The fundamental frequency associated to a faulty bit varies with
 the variations of $\dot{v}(t)$, hence the error in a faulty bit spreads
 across the frequency band when $\ddot{v}\neq0$.
\end{itemize}

\begin{figure}[h]
 \includegraphics[width=0.7\columnwidth]{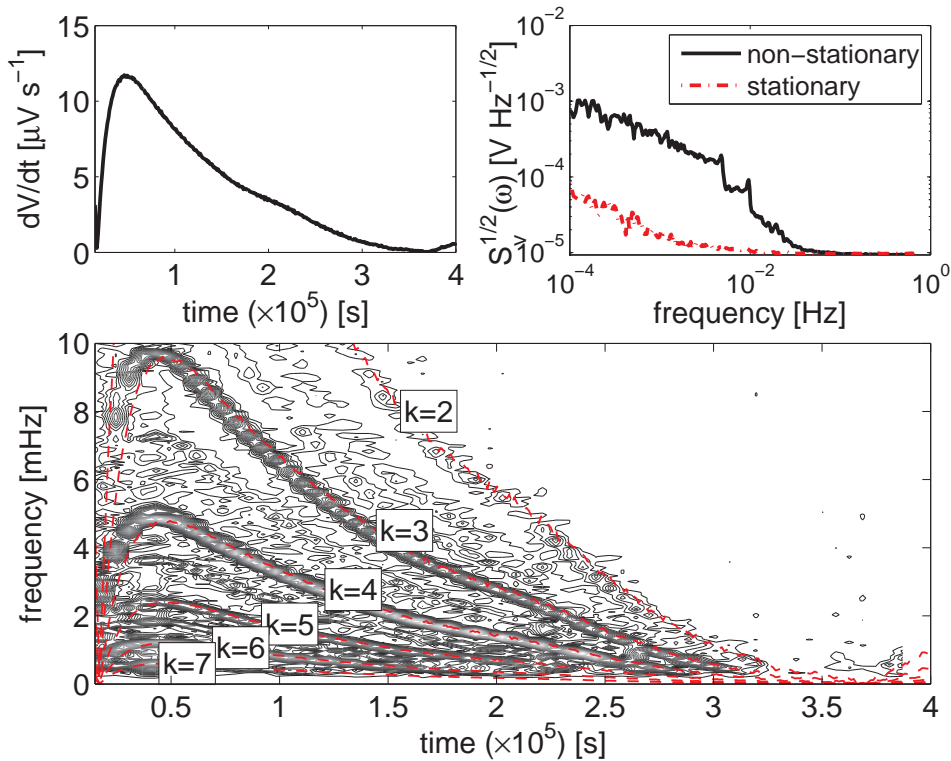}
 \caption{Top left: Absolute value of the input signal slope, $|\dot{v}(t)|$,
 of the measurement. It varies between 0 and 12.5\,$\mu$V\,s$^{-1}$. The
 sensitivity at the input of the ADC is $\simeq$1.35\,V\,K$^{-1}$, hence
 the values of the voltage slope almost directly provide the actual
 temperature drifts in [K\,s$^{-1}$]. Top right: power spectral density
 for: (i) the signal shown in the top left plot, i.e., a non-\emph{dc}
 signal (solid black trace) and, (ii) a \emph{dc} signal (dashed red trace),
 i.e., with $|\dot{v}(t)|\lesssim0.5$\,$\mu$V\,s$^{-1}$. The excess noise
 when measuring non-constant signals is very noticeable, and degrades the
 performance of the measurement by about one order of magnitude. Bottom:
 STFT (or spectrogram) of the measurement. The energy of the signal is
 concentrated in specific frequencies which change with time precisely
 following $|\dot{v}(t)|$, as predicted by Eq.(\ref{dw}). The latter is
 represented by dashed superimposed traces, and are labelled by the order
 of the corresponding faulty bit (from $k\/$\,=\,2 to $k\/$\,=\,7).
 Experimental results and theoretical estimates are in excellent agreement.
 \label{spec1}}
\end{figure}

The above can be validated by looking at real data from our LTP temperature
measurement system. We took a long time series of temperature data,
about 4$\times$10$^5$~seconds, and subdivided it into shorter stretches
of $\sim$16\,000~seconds. With this, we calculated the Short-Time
Fourier Transform (STFT)~\cite{oppenheim}, also known as \emph{spectrogram},
and plotted the results as shown in Figure~\ref{spec1} ---see caption for
details. The data are amenable to the analysis described above, and the
results indicate the presence of the foreseen frequency peaks associated
to faulty bits. They are clearly visible in Figure~\ref{spec1}, bottom
graph, as the darker areas, which show how the fundamental frequency of
the INL errors for different faulty bits is varying with time, and
accurately tracing the $|\dot v|$ curve shown on the top left plot.
A straightforward fit to the first equation~(\ref{dw}) permits the
identification of the corresponding faulty bit, thus confirming that
the extra noise in the temperature measurements is indeed due to the
INL errors of the ADC.

Once the problem has been identified, we focus on possible solutions to
fix it. The following section describes two different methods tested
successfully to avoid the consequences of INL errors in real ADCs.

\section{Mitigation of INL errors}
\label{mitigation}

Two different techniques have been used to deal with the non-linearities
of the ADC. The first is based on the injection of Gaussian noise out of
the MBW, taking advantage of the oversampling involved in the measurement.
The second is based on the addition of a triangular wave of high frequency
to the signal quantised by the ADC~\cite{gatti1963}. The following sections
describe both techniques as well as their practical implementation.

\subsection{Gaussian noise signal injection}
\label{dither.random}

In Section~\ref{dither.section} we saw that Gaussian noise can be used 
as a dither signal in order to mitigate the non-idealities of the ADC
transfer curve. We have also seen that the amount of noise generated
by the measurement system itself ($\sigma$$\simeq$$\Delta$) is not
enough to suppress the periodic components of erroneous bits for
$k\/$$\geq$2 ---see Fig.~\ref{spec1}. Thus, a natural solution is to
add more Gaussian noise to the ADC input. Obviously, this noise should
be added out of the MBW in order not to disturb the frequency range of 
interest, which is in the milli-Hertz range. The ADC sampling frequency
is 38.4\,kHz, thus leaving a large frequency slot to accommodate the
required additional noise. Care must however be taken to also place it
away from the fundamental frequency of the modulating square wave (and
its harmonics) so as to avoid bringing the Gaussian noise back into the
MBW in the demodulation process. The required amount of noise to be injected,
characterised by $\sigma$, basically depends on the input signal slope,
$\dot{v}(t)$, and on the frequencies of interest, and is limited by the
digital processing performed after quantisation ---see below.
Figure~\ref{slope1} shows the relationship between the input signal
slope and the main frequency component for each of the faulty bits. It
is useful to identify the faulty bits potentially affecting the measurement
as a function of the input signal drifts.
\begin{figure}[h]
\includegraphics[width=0.6\columnwidth]{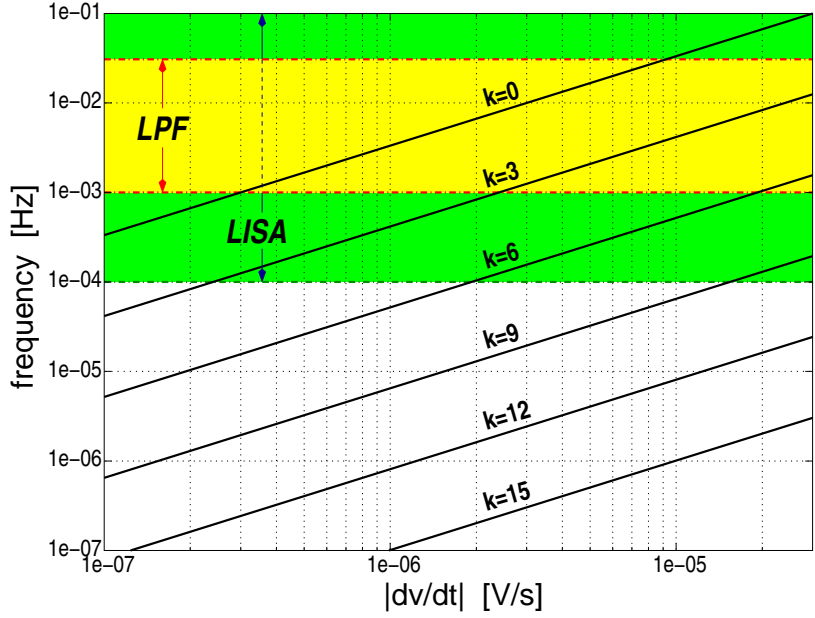}
 \caption{Relationship between the input signal slope, the faulty bit and the 
 fundamental frequency affected by the error in the $k\/$-th bit. When high
 input signal slopes are present, more faulty bits will affect the measurement
 quality in the MBW. Clearly, LSBs show up at higher frequencies than MSBs.
 The shaded areas span the MBWs of LISA (green) and LPF (yellow), as
 indicated. Note that LPF's MBW is a sub-region of LISA's. 
\label{slope1}}
\end{figure}

Once the problematic bits are identified, the needed amount of Gaussian
dither must be calculated. For this, we define the $\sigma$ needed to
filter out the noise components associated to the $k\/$-th faulty bit.
If we (conventionally) adopt a damping factor of 10 for the first harmonic
then $\sigma$ is easily derived from the condition (see Eq.~(\ref{qqq}))
\begin{equation}
 e^{-\xi^2\sigma^2/2}\leq\frac{1}{10}\ \ {\rm for}\ \ 
 \xi = \frac{2\pi}{2^{k+1}\,\Delta}
 \label{eq.14}
\end{equation}
whence
\begin{equation}
 \sigma\geq\frac{\sqrt{2\,\ln\,10}}{2\pi}\,2^{k+1}\,\Delta
 \label{sigma.req}
\end{equation}


\subsubsection*{Gaussian noise dither: practical implementation}
\label{hw.dither}

\begin{figure}[h]
 \includegraphics[width=0.75\columnwidth]{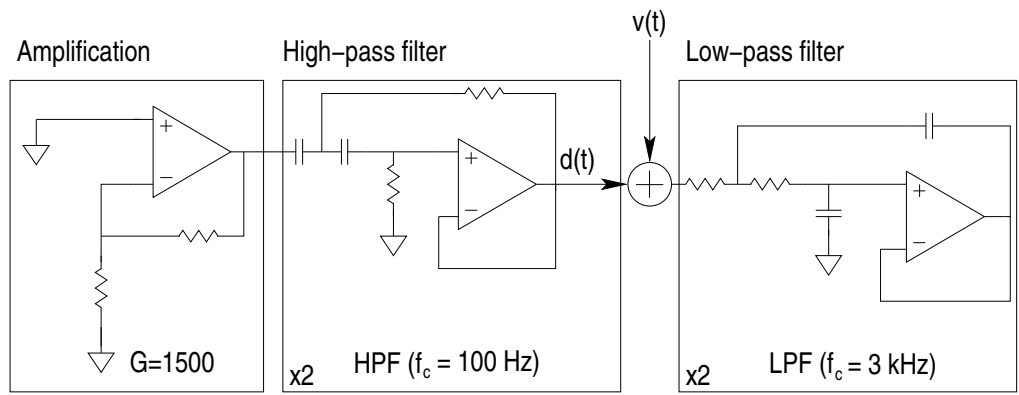}
 \caption{Gaussian noise dither signal generator. Its power spectral
 density is shown in Fig.~\ref{noisegen}. $v(t)$ is the amplified signal
 from the Wheatstone bridge ---see Figure~\ref{fee.proc}. $d(t)$ is the
 dither signal before low-pass filtered.
 \label{noisegen_circuit}}
\end{figure}

\begin{figure}[h]
 \includegraphics[width=0.5\columnwidth]{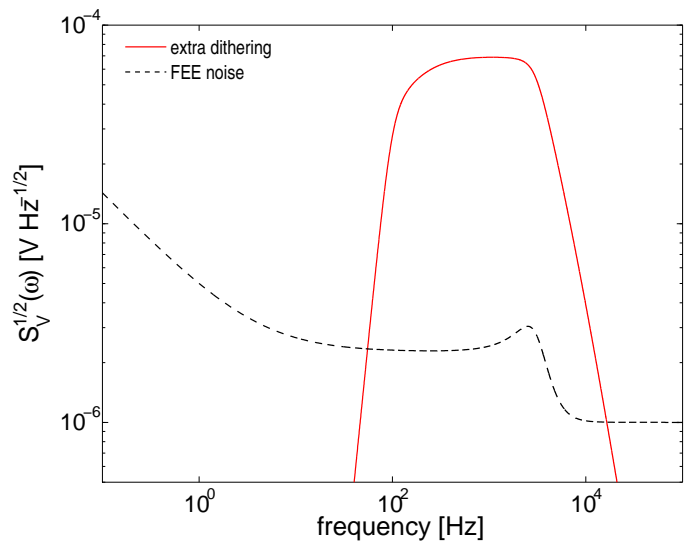}
 \caption{Solid red trace: Noise shape of the dither signal generated by
 the circuit shown in Figure~\ref{noisegen_circuit}. The central plateau
 corresponds to $\sigma$\,=\,3\,mV ---see Section~\ref{T.R}.  The dashed
 black trace is the floor noise of the measurement system itself, prior
 to demodulation~\cite{fee.paper}.
 \label{noisegen}}
\end{figure}

In this section we briefly describe the hardware implementation
of the Gaussian dithering scheme. The circuit diagram is given in
Fig.~\ref{noisegen_circuit}: the first stage amplifies the noise
of an operational amplifier (OP-07~\cite{analog.devices}); the second
stage is a high-pass filter of~4-th order (two Sallen-Key filters in
cascade) with a cut-off frequency of~100\,Hz; the third stage is an
adder that sums the signal of interest (the amplified output of the
Wheatstone bridge) to the dither signal. In the fourth stage, the sum
of signal and dither are low-pass filtered with a~4-th order low-pass
filter (again, two Sallen-Key filters in cascade) with cut-off
frequency~3\,kHz. The output of this chain is fed to a~16-bit SAR ADC.
The noise shape of the dither signal is given in Fig.~\ref{noisegen}
(solid trace) where the characteristic frequencies of the high- and
low-pass filters are clearly visible.

As mentioned in Section~\ref{fee.description}, the amount of Gaussian
dither must be kept under control to avoid excessive noise folding back
into the MBW during the digital demodulation stage. The limit of the
usable Gaussian noise amplitude, $\sigma$, can be easily estimated
assuming a flat spectrum, $S_{V, \rm \, dither}$, from $\omega_{\rm m}/2\pi$
(the corner frequency of the high-pass filter) to $\omega_{\rm M}/2\pi$
(the corner frequency of the low-pass filter). The demodulation by a
square wave of frequency $\omega_{\rm c}$ entails a certain gain at
\emph{dc} of the odd harmonics of $\omega_{\rm c}$ of the dither noise
which, under the just mentioned assumption, results in the
following~\cite{oppenheim}:
\begin{equation}
 S_{V,\,{\rm extra}}(\omega=0)\simeq S_{V, \rm \,dither}\cdot
 \frac{4}{\pi^{2}}\;\sum_{n=\left\lfloor\omega_{\rm m}/2\omega_{\rm c}\right\rfloor}%
 ^{\left\lfloor (\omega_{\rm M}-\omega_{\rm m})2\omega_{\rm c}\right\rfloor}\;
 \frac{1}{(2n+1)^2}
 \label{sdc}
\end{equation}
where $\lfloor \ \rfloor$ is the \emph{floor function}, e.g.,
$\lfloor x\rfloor$ is the largest integer smaller than $x$.

In order to cut down the noise leaking into the MBW, i.e., at \emph{dc}
in practice, we again impose a requirement that it be less than 10\% of
the floor noise in the absence of ADC errors, $S_{V,\,{\rm FEE}}$, i.e.,
\begin{equation}
 S_{V,\ {\rm extra}}(\omega=0)\leq\frac{S_{V,\rm \, FEE }(\omega)}{10}
 \label{eq.15}
\end{equation}

In our case, $\omega_{\rm m}/2\pi\simeq 100$\,Hz,
$\omega_{\rm M}/2\pi\simeq 3$\,kHz, and $\omega_{\rm c}/2\pi=6.25$\,Hz,
so that the $\Sigma$-sum in Eq.~(\ref{sdc}) is approximately 0.015, hence
$S_V(\omega\!=\!0)$\,$\simeq$\,0.015\,$S_{V,\,\rm dither}$. The value
of $S_{V,\,{\rm FEE}}$ is $\simeq 1.5\times10^{-11}$\,V$^2$\,Hz$^{-1}$ in the MBW.
The maximum value for $S_{V, \rm \,dither}$ compliant with the criterion
Eq.~(\ref{eq.15}) is therefore $10^{-10}$\,V$^{2}$\,Hz$^{-1}$, which
corresponds to $\sigma\simeq$\,0.5 mV.

In Section~\ref{results}, we present the tests and results obtained with
this technique.

\subsection{Triangular wave dither} 
\label{dither.det}

Another implemented and tested technique consists in using a deterministic
signal instead of random Gaussian noise as the dither signal. Different
signals can be used for this purpose, but for simplicity we have used a
triangular wave form in the generation of the signal ---see below. In this
section we describe the theoretical basis and the hardware implementation
details.

The dither signal is now a triangular wave, which can be represented
by the Fourier series
\begin{equation}
 d(t)=\frac{D_{\rm o}}{2}-\frac{4D_{\rm o}}{\pi^2}\sum^{\infty}_{n=0}\,
 \frac{1}{(2n+1)^2}\,\cos{(2n+1)\omega_{\rm tr} t}
\end{equation}
where $D_{\rm o}$ and $\omega_{\rm tr}$ are the amplitude and the frequency
of the triangular wave, respectively ---see Fig.~\ref{triangular.scheme}.
Its pdf is~\cite{papoulis}
\begin{equation}
 p(d)=
 \begin{cases}
 \frac{1}{D_{\rm o}} & 0\leq d\leq D_{\rm o} \\[0.5ex]
 0 & d<0 \ {\rm and} \ d>D_{\rm o}
 \end{cases}
\end{equation}
whose Fourier transform is
\begin{equation}
 {\cal P}(\xi) = \frac{\sin (D_{\rm o}/2)\xi}{(D_{\rm o}/2)\xi}
\end{equation}

As shown in Section~\ref{problem}, the averaged quantisation error is
\begin{equation}
 \langle {\cal Q}(\xi)\rangle = {\cal Q}(\xi){\cal P}^*(\xi)
 \label{dith.triang.eq}
\end{equation}

\begin{figure}[h]
 \includegraphics[width=0.7\columnwidth]{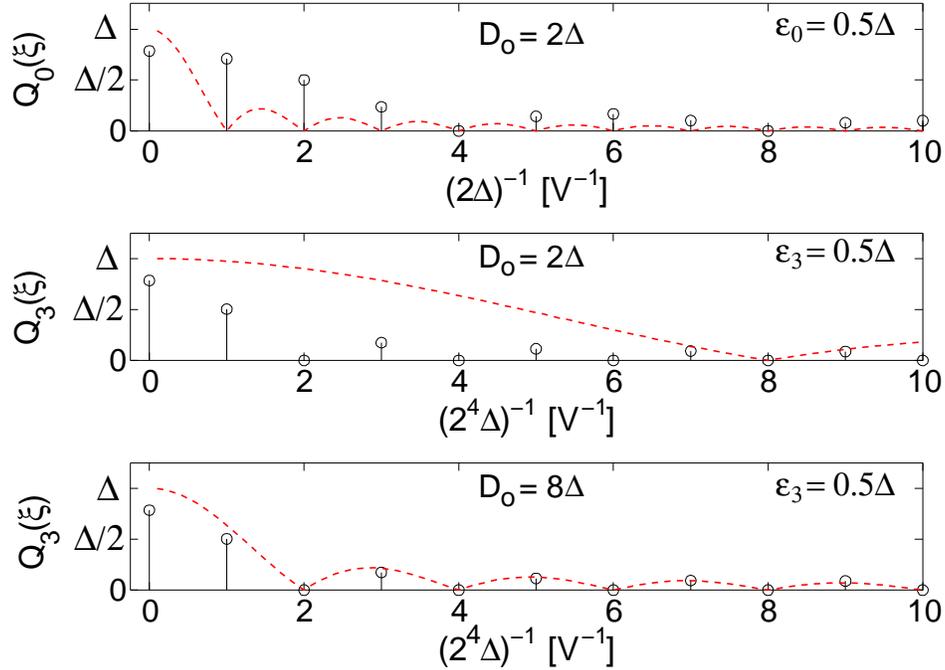}
 \caption{Top: quantisation error for a real ADC with a faulty bit in
 $k\/$=0, $\epsilon_0=0.5\Delta$, and the low-pass filter when using
 a triangular wave as the dither signal with $D_{\rm o}=2\Delta$.
 Centre: same as above for an error in bit $k\/$=3, $\epsilon_3=0.5\Delta$.
 Bottom: same as above but with $D_{\rm o}=8\Delta$ instead of $2\Delta$.
 Note that $x$-axis scales are different for ${\cal Q}_0$ and
 ${\cal Q}_3$.
 \label{triangular.dither}}
\end{figure}

Eq.~(\ref{dith.triang.eq}) is graphically evaluated in
Figure~\ref{triangular.dither} for different faulty bits
and triangular wave amplitudes.

Like we did with the Gaussian noise, we can give an expression to determine
the required triangular wave amplitude, $D_{\rm o}$, to attenuate the
effect of the $k\/$-th bit:
\begin{equation}
 \left|\frac{\sin{(D_{\rm o}/2)\xi}}{(D_{\rm o}/2)\xi}\right|
 \leq\frac{1}{10}
 \quad {\rm for}\quad \xi = \frac{2\pi}{2^{k+1}\,\Delta}
 \label{sinx.x}
\end{equation}

There is no closed form solution to Eq.~(\ref{sinx.x}) for the amplitude
of the triangular wave, which requires numerical evaluation in each specific
case. The amplitude $D_{\rm o}$ needed to attenuate the errors coming from
the $k\/$=6 bit is $\sim$20\,mV.

\subsubsection*{Triangular wave dither: practical implementation}
\label{triangle.hw}

\begin{figure}[h]
 \includegraphics[width=0.45\columnwidth]{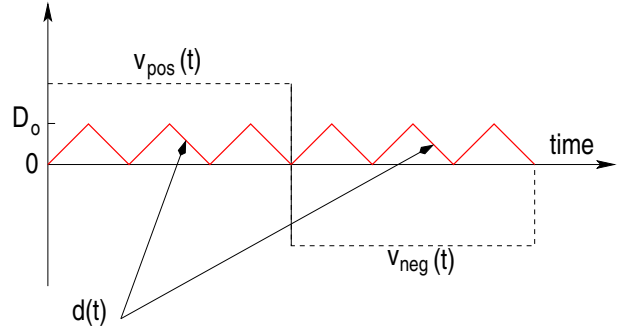}
 \caption{Triangular wave dither signal and signal coming from the 
 measurement chain (dashed trace).
 \label{triangular.scheme}}
\end{figure}

Figure~\ref{triangular.scheme} shows the triangular wave added to the system.
It is important to note that the addition of the triangular signal will not
perturb the temperature measurement within the MBW. The digital demodulation
involved in the measurement ---see Sec.~\ref{fee.description}--- is done
by averaging 6144 samples: 3072 during one polarity and 3072 during the
opposite polarity. Afterwards they are subtracted and divided by 2. Thus,
if we inject exactly the same signal in both polarities the net contribution
of the added dither signal to the output is zero ---see Eq.~(\ref{vo}).

The output signal after the digital processing is
\begin{equation}
 v_{\rm o} = \frac{[\bar{v}_{\rm pos}(t)+\bar{d}(t)]-[\bar{v}_{\rm neg}(t)
 + \bar{d}(t)]}{2} =
 \frac{\bar{v}_{\rm pos}-\bar{v}_{\rm neg}}{2}
 \label{vo}
\end{equation}
where $d(t)$ is the triangular wave ---see Fig.~\ref{triangular.scheme}---
and an overbar (\ $\bar{}$\ ) means average over 3072 samples.

Eq.~(\ref{vo}) indicates that there is no limit on the amplitude of the
triangular dither, provided triangular waves in both polarities are
identical. Therefore that non-linearities of MSBs can be reduced without
degrading the measurement. This happens for an analog dither signal, but
ours is actually generated with a DAC, which imposes some limits on the
validity of the previous statement ---see below.

\begin{figure}[h]
 \includegraphics[width=0.6\columnwidth]{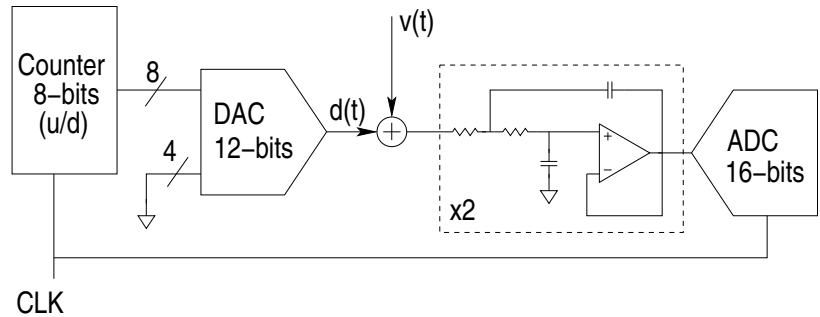}
 \caption{Triangular wave dither signal generator. $v(t)$ is the amplified
 output signal of the Wheatstone bridge. $d(t)$ is the triangular wave.
 \label{triangular.circuit}}
\end{figure}

The circuit which implements the triangular signal is
shown in Figure~\ref{triangular.circuit}, and consists in an 8-bit
up-and-down counter followed by a 12-bit DAC where only the 8 LSBs
are used. The circuit is configured such that the quantisation step
of the DAC closely matches that of the 16-bit ADC. Mismatches here
result in actual performance diverging from the predictions of the
theoretical analysis described above.

The output signal from the DAC is added to the signal of
interest and both are low-pass filtered (a 4-th order Sallen-Key filter
with a cut-off frequency of 3\,kHz), then quantised by the 16-bit SAR ADC.
The triangular wave is low-pass filtered to eliminate high-frequency 
components related to the digital quantisation of the DAC. Nevertheless,
the previous analysis is still valid since the fundamental frequency of
the triangular wave is 50\,Hz and the low-pass filter cut-off frequency
is 3\,kHz. Thus, the dither signal will go almost unaltered through the filter
(except for some distortion in its high-frequency components).

\section{Test set-up and results}
\label{results}

Both methods described in Section~\ref{mitigation} have been put to test.
For comparison, a 24-bit Delta-Sigma ADC (LTC2440~\cite{linear.tech})
which, in principle, should exhibit less non-linearity problems, has also
been tested. The two dither techniques under study have been tested
using exactly the same electronics, the same 16-bit SAR ADC
AD977~\cite{analog.devices}\footnote{
Actually, the ADC used for the space model is the Texas Instruments
ADS7809 which is based in the same structure.},
and under the same input signal conditions. In this section first we
give a brief description of the test set-up and, then, the obtained
results are presented.

\subsection{Test set-up}

\begin{figure}[h]
 \includegraphics[width=0.6\columnwidth]{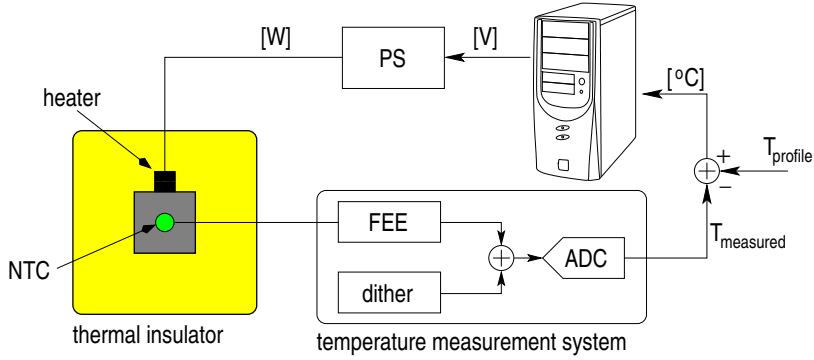}
 \caption{Setup scheme. FEE labels the analog signal processing of the
 temperature measurement system and PS stands for the programmable power
 supply.
 \label{setup.fig}}
\end{figure}

\begin{figure}[h!]
 \includegraphics[width=0.5\columnwidth]{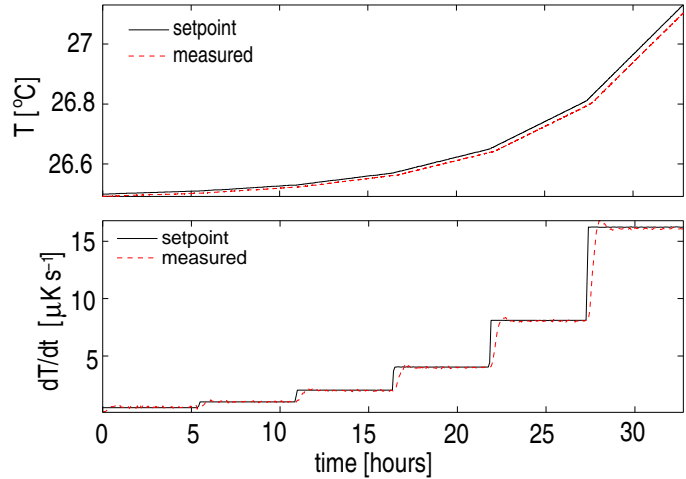}
 \caption{Top: nominal and experimental (achieved with the feedback
 temperature control shown in Fig.~\ref{setup.fig}) temperature profiles.
 Bottom: time derivative of the signals shown in the top plot. Tested
 slopes are: 0.5, 1, 2, 4, 8 and 16\,$\mu$K\,s$^{-1}$.
\label{profile.experimental}}
\end{figure}

The test set-up is composed by different parts. Temperature sensors are
placed inside a thermal insulator designed to screen out ambient temperature
fluctuations to the required level in the MBW, i.e.,
$S_T^{1/2}(\omega)\lesssim 10$\,$\mu$K\,Hz$^{-1/2}$ for
$\omega/2\pi\gtrsim1$\,mHz~\cite{on.ground.tests}. A temperature control is
included to implement different temperature profiles inside the insulator;
basically, a set of temperature ramps with different slopes is generated
in order to assess whether or not the methods to overcome the INL errors
of the ADC work ---see Figure~\ref{profile.experimental}. The temperature
control consists in a heater commanded by a programmable power supply which
is in turn controlled by the computer calculated value of the difference
between the desired temperature and the actual measurement ---see
Figure~\ref{setup.fig}.

The nominal and actual profiles are shown in
Figure~\ref{profile.experimental}. The experiment has been repeated for
the two dithering techniques, and for the 24-bit Delta-Sigma ADC as well.
Prior to that, the 16-bit SAR ADC with no dither signal was tested in
order to provide a reference measurement with the INL effects in it.

\subsection{Experimental results}
\label{T.R}

\begin{figure}[h]
 \includegraphics[width=0.83\columnwidth]{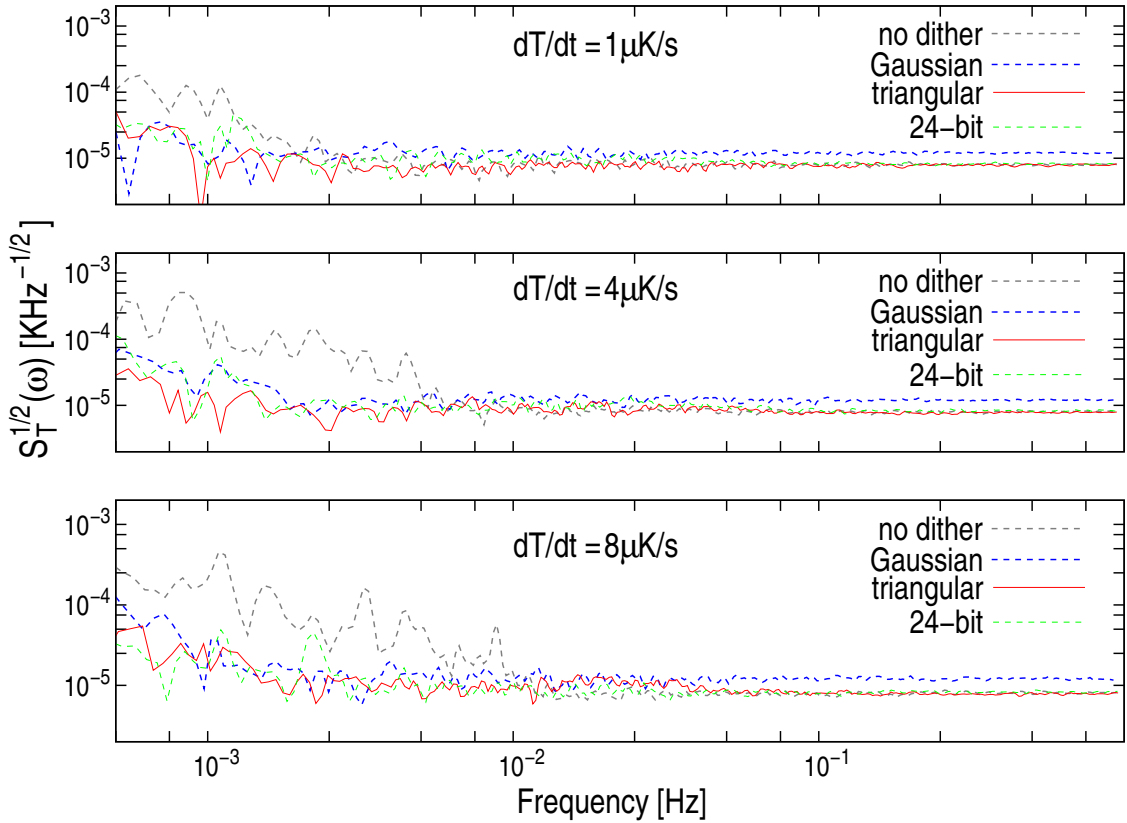}
 \caption{Linear Power Spectral Density for different configurations:
 16-bit ADC without dither, 16-bit ADC with Gaussian dither, 16-bit
 ADC with triangular wave dither, and 24-bit Sigma-Delta ADC without
 dither. Input signals are ramps with slopes varying from 1 to
 8\,$\mu$K\,s$^{-1}$. In absence of dither, the noise of the ADC increases
 (in amplitude and bandwidth) with the slope of the input signal. The
 solutions adopted work satisfactorily for slopes up to
 $\sim$8\,$\mu$K\,s$^{-1}$ level. The 24-bit Sigma-Delta does not exhibit
 non-linearity problems for the tested slopes. See text in Section~\ref{T.R}
 for details.
 \label{plot.paper}}
\end{figure}

The power spectral density in the different tested configurations
(no dither, Gaussian noise, triangular wave and 24-bit Delta-Sigma ADC)
are given in Fig.~\ref{plot.paper} for three different slopes, more
specifically, for 1, 4 and 8\,$\mu$K\,s$^{-1}$.
The measurements performed with the 16-bit SAR ADC are clearly affected 
by the INL errors when no dither signal is used. The noise in the MBW
increases by more than one order of magnitude when slopes are around
1\,$\mu$K\,s$^{-1}$ and above. Figure~\ref{plot.paper} shows how the INL
errors appear in the MBW, spanning wider frequency regions as slopes
increase, thus confirming the predicted behaviour. For instance,
when the slope is 4\,$\mu$K\,s$^{-1}$ the INL effect is only noticeable
at frequencies below 6\,mHz, while for a slope of 8\,$\mu$K\,s$^{-1}$
the noise appears at frequencies as high as 10\,mHz.

When the dither signal is Gaussian noise, the INL effects can be
satisfactorily reduced down to 1\,mHz for drifts under~8\,$\mu$K\,s$^{-1}$.
Looking up Figure~\ref{slope1}, we see that faulty bits up to $k\/$\,=\,5
will create additional noise in the band. Then, using Eq.~(\ref{sigma.req}),
we find that the necessary dither requires $\sigma$\,$\simeq$\,22\,$\Delta$,
or $\sigma$\,$\simeq$\,3\,mV. This is however a factor of 6 larger than
the maximum estimated after Eq.~(\ref{eq.15}), which means some extra
noise will be added to the system floor noise, if we insist on applying
a $\sigma$\,=\,3\,mV dither, as discussed in Section~\ref{hw.dither}.
We did take this option, with the result that the floor noise becomes
a factor $\sim$1.5 larger than nominal [in good agreement with predictions
calculated using Eq.~(\ref{sdc})], with the advantage that good damping
of the ADC's INL errors obtains.

\begin{figure}[h!]
 \includegraphics[width=0.5\columnwidth]{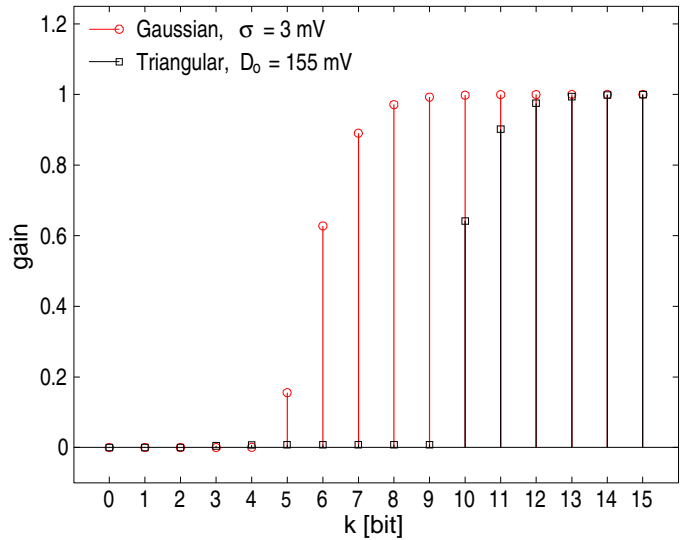}
 \caption{Gain of the equivalent filters of the Gaussian noise dither and the
 triangle wave. The gain shown for each bit corresponds to the fundamental
 frequency. Gaussian noise dither mitigates the  error in the bits
 $k\/$\,$\leq$\,5 whereas the triangular wave attenuates the
 error in the bits $k\/$\,$\leq$\,9.
 \label{plot.B1}}
\end{figure}

The use of a triangular wave as the dither signal appears as the most
robust option to deal with the non-idealities of the ADC: on the one
hand, the floor noise is left untouched and, on the other hand, immunity
to faulty bits and high signal slopes can be tuned essentially at will.
The 24-bit Delta-Sigma ADC exhibits a behaviour similar to that observed
in the measurements performed by the 16-bit ADC with added dither signals.
Non-linearity errors in this ADC are not noticeable when measuring signals
drifting up to~8\,$\mu$K\,s$^{-1}$. The generated triangular wave had the
following properties: $D_{\rm o}$=155\,mV and a period of 20\,ms, which is
an integer sub-multiple of the duration of each polarity (80\,ms)
---see Figure~\ref{triangular.scheme}. This amplitude is almost 8 times
higher than the one estimated towards the end of Section~\ref{dither.det}
(20\,mV) to attenuate errors up the $k\/$\,=\,6~bit. This means that
immunity to faulty bits under higher temperature drifts is accomplished.
Figure~\ref{plot.B1} provides a clear display of the superiority of
triangular wave over Gaussian noise dither.

Summing up, Figure~\ref{plot.paper} confirms the analysis in this paper,
and shows that when slopes of $\sim$10\,$\mu$K\,s$^{-1}$ are present, the
effects of the non-idealities can be made negligible by use of proper
dither. For higher slopes, e.g., 16\,$\mu$K\,s$^{-1}$, some increase
in the power spectrum near one milli-Hertz is detected. We believe
this is probably due to the relative simplicity of the model used.
Further research is ongoing to clarify these matters. Nevertheless,
such drifts look unlikely both in LISA and LPF.

\section{Discussion}
\label{discussion}

Temperature diagnostics measurements in LISA and LISA PathFinder must meet
very demanding requirements: very low noise, a few $\mu$K\,Hz$^{-1/2}$, and
very low frequency band, below 1 Hz and down to fractions of a milli-Hz.
After a suitable measurement system was in place, extra noise was seen
to appear at the lower end of the MBW when slowly drifting temperatures
were measured. For temperature slopes higher than
$\sim$0.5\,$\mu$K\,Hz$^{-1/2}$ this extra noise challenges the performance
of the LTP measuring system, the problem growing more severe at lower
frequencies, down into the submilli-Hertz LISA band.

The source of the noise has been investigated and identified as due to
the INL effects of the ADC, which is a 16-bit SAR ADC\,\footnote{
Space qualification constraints prevent use of more precise ADCs.}.
We have laid down the theoretical basis of the problem, which has been
validated by laboratory experiment.

Once the problem is well understood, it needs to be solved. Two different
options to mitigate this effect have been proposed, analysed and tested.
Both are based on the addition of a dither signal to the signal of
interest prior to the ADC quantisation. In one of them, we use Gaussian
noise, which certainly reduces the INL error effects of the ADC, although
in our case its performance is limited by the the digital processing which
sets an upper limit on the noise amplitude, $\sigma$, which can be injected
in the measurement. Alternatively, dithering with a triangular wave does
not add noise in the MBW, and generally results in a more efficient attenuation
of INL errors. A 24-bit Delta-Sigma ADC has been used for comparison, with
the result that a 16-bit ADC plus averaging (to increase the resolution)
and proper dither reaches the same performance as a 24-bit Delta-Sigma ADC.

Summing up, a robust method to suppress the INL errors of the ADC
in the temperature measurement system of the LPF mission has been presented
in this paper. Such a system is not foreseen to fly in LPF, but the results
obtained herein are important in view of the LISA mission which will need
similar (or even higher) precisions at lower frequencies. The method we
have described is also useful for other subsystems of LISA where ADC
non-linearities appear as a limitation in the performance of the measurement
system.

\begin{acknowledgments}

Support for this work came from Project ESP2007--61712 of Plan Nacional
del Espacio of the Spanish Ministry of Education and Science (MEC).
JS acknowledges a grant from MEC.
\end{acknowledgments}

\end{document}